# Spatiotemporal Dynamics of Self-Similar Parabolic Pulse Evolution in Multimode Fibers

Leila Graini[1, 2] and Bülend Ortaç[2], *Member, IEEE*

[1]Telecommunications Laboratory, 8 Mai 1945-Guelma University, Guelma 24000, Algeria
[2]National Nanotechnology Research Center and Institute of Materials Science and Nanotechnology, Bilkent University, Ankara 06800, Turkey (e-mail: graini.leila@univ-guelma.dz, ortac@bilkent.unam.edu.tr .

*Abstract*— In this paper, we investigate for the first time, the spatiotemporal dynamics of self-similar parabolic pulse evolution in multimode fibers (MMFs). We present numerical predictions of the existence of parabolic pulse in a graded and step index MMFs and demonstrate that the parabolic pulse formation process could have a prominent effect on the spatiotemporal behavior in both fibers. We study the effect of the input pulse energy and its initial modal composition on the resulting parabolic pulse and the corresponding spatial beam profile. When the fundamental mode is mostly excited, the pulse evolves into a linearly chirped pulse with parabolic intensity shape that propagates self-similarly in both fibers. This allows for efficient and high-quality pulse compression with sub-40 *fs* pulse duration. Dependence on input pulse shapes, its chirp parameter and sign towards conversion into parabolic shape are also reported. We quantify the reshaping of the parabolic pulse using the misfit parameter. We demonstrate a route to the parabolic pulse evolutions for all modes equal energy distributed initial condition. In addition, we also observe that the beam profiles of output fields could be different for same MMF with different initial pulse energy. Moreover, a spatiotemporal nonlinear dynamic, auto-selection beam of one specific mode is investigated. Thus remain spatio-temporally stable for more than a decade of the input pulse energy variation. Parabolic pulse formation process plays critical roles in these nonlinear dynamics. This approach provides another framework to understand the complex nonlinear dynamics in MMFs.

## I. Introduction

Self-similar parabolic pulse generation in single mode fibers (SMFs), first predicted in 1993 by Anderson et al [1] and then demonstrated in 2000 by Fermann et al [2], has been widely investigated and is still an active area of research due to its unique characteristics and numerous applications for potential scaling approach of fiber laser systems to unprecedented pulse energy and power levels [3-12]. Such pulse is an attractor solution to the nonlinear Schrödinger equation (NLSE) in normal dispersion regime with the presence of nonlinearity. It is also present robustness to the wave-breaking under high intensity condition, as well as, its characteristic linear chirp can be lead to efficient pulse compression. Therefore, parabolic pulse has considered as candidates for making fiber amplifiers and generating highly energetic short pulses. However, SMFs are challenged by the ever-increasing demand for higher-energy pulses. The use of MMFs represents an important potential platform for overcoming this limitation due to the corresponding larger mode area and the additional spatial degree of freedom compared to SMFs.

Recently, spatiotemporal pulse propagation dynamics in MMFs have attracted huge attention as a testbed to investigate uncovered nonlinear effects. Among these effects, spatiotemporal optical solitons, supercontinuum generation, spatiotemporal instability, self-similar fiber laser and mode-locking [13-17] have been proposed. However, only one study has investigated parabolic pulse generation in MMF [18]. The fiber used in this investigation is step index fiber type and is excrementally tapered allowing single mode fiber properties. Furthermore, the multimode nonlinear dynamics is not considering and the theoretical analysis is based on the (1 +1)-D NLSE. Generally speaking, at present, there is no research work carried out for (1 + 3)-D NLSE to generate such pulse in MMFs. Then parabolic pulse generation and its spatiotemporal dynamics in multiple modes fiber structures will be a question to investigate.

In this paper, we investigate that spatiotemporal parabolic pulse can be efficiently generated in different MMFs under various input conditions. We also analyze its spatial behavior during propagation through two different passive MMFs. Graded and Step-index multimode fiber (GRIN-F and STEP-F) platforms are good candidate for understanding the complex spatiotemporal nonlinear dynamics of parabolic pulses. In order to study the dynamics of parabolic pulses in MMFs compared by those generated in SMFs, we have first started by the condition of the predominantly excitation of the fundamental mode. We show that it is possible to achieve a well-matched parabolic pulse shapes in both MMFs with different mechanisms of the propagations due to the MMFs properties. The characteristics of the self-similar evolution and parabolic pulse shape with linear chirp is also provided in both MMFs. Compressions of the output parabolic pulses lead to the generation of sub-40 *fs* pulses. In contrary to the similariton, parabolic pulse shape in passive fibers strongly depends on the initial pulse parameters [5, 7]. Parabolic pulses formed from different input pulse shapes such as Gaussian, secant hyperbolic, super-Gaussian pulses are also studied as well as the influence of the initial chirp on the parabolic pulse formation. The transverse distribution intensity of the beams remains Gaussian-like profile along the both MMFs in the case when the fundamental mode is predominantly excited. Our second attention is notably in the spatiotemporal behavior of parabolic pulses under the condition of the equal energy distributed among all the modes. We studied the effect of the input pulse energy and its initial modal composition on the resulting parabolic pulse and its corresponding spatial beam profile. The results showed that all the modes evolved to the parabolic shape in both MMFs. In addition, parabolic pulse formation process remains spatial beam profile stable for more than a decade of its initial energy. Moreover, we observed an interesting nonlinear phenomenon of beam auto-selection represented by the evolution of the multimode spatial beam profile into a single mode profile.

## II. NUMERICAL MODELING FOR PARABOLIC PULSE GENERATION IN MMFs

Simulations of spatiotemporal parabolic pulse evolution in MMFs are performed using (NLSE) in (3+1)-D [19], which is solved by the fourth-order Runge–Kutta method [20].The (3+1)-D NLSE (1) is including the effects of stimulated Raman scattering, dispersion up to forth order, self-steepening and intermodal effects. The different coefficients are calculated from the numerically-calculated modes for a MMFs based on the one used in [21].

$$\partial_z A_p(z,t) = i\left(\beta_0^{(p)} - \Re[\beta_0^{(0)}]\right)A_p - \left(\beta_1^{(p)} - \Re[\beta_1^{(0)}]\right)\frac{\partial A_p}{\partial t} + i\sum_{n\geq 2}\frac{\beta_n^{(p)}}{n!}\left(i\frac{\partial}{\partial t}\right)^n A_p + i\frac{n_2\omega_0}{c}\left(1 + \frac{i}{\omega_0}\partial t\right)\sum_{l,m,n}\left\{(1-f_R)S_{plmn}^k A_l A_m A_n^* + f_R A_l S_{plmn}^R \int_{-\infty}^t d\tau A_m(z,t-\tau)A_n^*(z,t-\tau)h_R(\tau)\right\} \quad (1)$$

Where $A_p$ (z, t) is the electric field of mode p. $S_{plmn}^k$ and $S_{plmn}^R$ are the nonlinear coupling coefficients for the Kerr and Raman effects, respectively, with $p$, $l$, $m$, and $n$ representing the numbers of spatial modes, $f_R$ is the fractional contribution of the Raman effect ($f_R = 0.18$), $h_R$ is the delayed Raman response function, and $\beta_n^{(p)}$ are the higher-order dispersion coefficients of mode P. $\Re[...]$ denotes the real part only and $n_2$ is the nonlinear index of refraction ($3.2 \times 10^{-20}$ m$^2$W$^{-1}$).

The MMFs under consideration are assumed to have a core radius of 25 μm, a parabolic index and step reflective index profiles in the GRIN-F and the STEP-F with index contrast and numerical aperture of Δ=0.0068 and $NA = 0.17$, respectively. These MMFs are excited at the wavelength of 1030 nm where the dispersion is normal. The MMFs characterizations are summarized in the Tab.1.We considered the first six linearly polarized modes (LP$_{01}$, LP$_{11a}$, LP$_{11b}$, LP$_{02}$, LP$_{21a}$, and LP$_{21b}$).

TABLE I
USED PARAMETERS FOR TWO MMFs

| MMF | Effective Area (A$_{eff}$) in um$^2$ | Nonlinear coefficient (γ) in W$^{-1}$km$^{-1}$ | Normal dispersion coefficient ($\beta_2$) in ps$^2$km$^{-1}$ LP$_{01}$, LP$_{11a}$, LP$_{11b}$, LP$_{21a}$, LP$_{21b}$, LP$_{02}$ |
|---|---|---|---|
| GRIN-F | 150.67 | 1.29 | 18.96, 18.95, 18.95, 18.94, 18.94, 18.94 |
| STEP-F | 963.84 | 0.20 | 18.71, 18.32, 18.32, 17.80, 17.80, 17.62 |

In order to check the quality of the formed parabolic pulse, we using misfit parameter (M) between the pulse temporal intensity profiles A and a parabolic fit A$_p$ of the same energy [5]:

$$M^2 = \frac{\int |A|^2 - |A_p|^2 d\tau}{\int |A|^4 d\tau} \quad (2)$$

The expression for a parabolic pulse of energy $U_p = 4P_p T_p/3\sqrt{2}$ is given by:

$$\begin{cases} A_p(T) = \sqrt{P_p}\sqrt{1 - 2T^2/T_p^2} & |T| \leq T_p/\sqrt{2} \\ A_p(T) = 0, & |T| > T_p/\sqrt{2} \end{cases} \quad (3)$$

Where $P_p$ is the peak power of the parabolic pulse and $T_p$ is the time duration. The less value of M shows better fit to the parabolic waveform. We consider that the pulse shape is parabolic for M≤0.04 [5].

## III. PARABOLIC PULSE GENERATION THROUGH THE PROPOSED GRIN-F

### A. Under the condition of the predominant fundamental mode excitation

In this simulation, we assume a Gaussian-shaped input pulse with a pulse duration of 200 *fs* propagates in 1m of the proposed GRIN-F. The input pulse energy is considered to be 7 nJ corresponding the input pulse peak power of about 35 kW. We set this energy mostly coupled in the LP$_{01}$ mode as 99 %. The energy and the coupling condition for the parabolic pulse generation are well chosen to ensuring the formation of parabolic pulse and avoid the energy transfer into the other modes at the first stage. The characteristic nonlinear length in this case is 22.15 mm which is higher than the minimum walk-off length of 7.5 mm. The pulse evolution is shown in Fig.1, which shows the explicitly of the characteristics of the self-similar regime in both the temporal and spectral domains. The temporal and the spectral widths increase exponentially and the amplitudes decrease with propagation inside the proposed GRIN-F and it could be clearly observed the absence of any pulse deformation or oscillation causing by wave breaking phenomena [1].We know that the pulse reshaping inside the GRIN-F is governed by the interplay between the nonlinearity presented by self-phase modulation (SPM) and normal dispersion, two propagation regimes are observed, alike to that found in previously studies [2, 3, 8]. The first regime implies reshaping of the pulse in a short propagation distance determining a transient state of the pulse evolution in the GRIN-F. In this part, the strong action of SPM effect leads to fast changes of the pulse shape and its spectrum, with a nonlinear pulse chirp. In the second regime, the pulse profile converges to the parabolic shape and evolves in a self-similar manner. In this part, dispersion is predominant than the SPM, which appeared by the large broadening of the pulse duration and the slow variation of the spectrum expansion witch approaches its maximal, as well as the chirp becomes linear.

Fig.2 shows the pulse temporal and spectral shapes and the chirp profile at the output of the GRIN-F. It can be seen that after propagation in 1 m of the fiber, the LP$_{01}$ mode evolves towards a parabolic shape pulse with a quasi-linear chirp in the entire duration of the pulse. This behavior is also good agreement with theoretical predictions between the temporal pulse shape and the parabolic fit (black dashed line). The generated parabolic pulse duration is about 2 ps. In order to quantify the quality of the generated parabolic pulse, misfit parameter (M) is calculated using equation (2) and plotted in Fig. 2.c versus the distance of the propagation for the temporal pulse shape. We can see that M is minimum value of 0.013 at propagation distance of 1m and the pulse shape indeed is very close to the ideal parabolic waveform. The spectral evolution is also shown in Fig.2.b and a substantial broadening of the spectrum with 60 nm spectral width is obtained. The above results imply that the parabolic pulse generated in the GRIN-F can be compressed down to 40 *fs* as presented in Fig.2.d, so that compression factor of 5 is obtained.

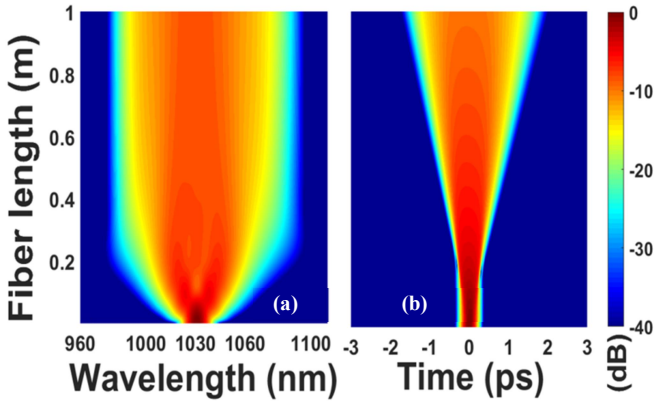

Fig.1 Self-similar evolution of parabolic pulse for spectral (a) and temporal (b) domains in GRIN-F where the most initial energy is coupled in the $LP_{01}$ mode.

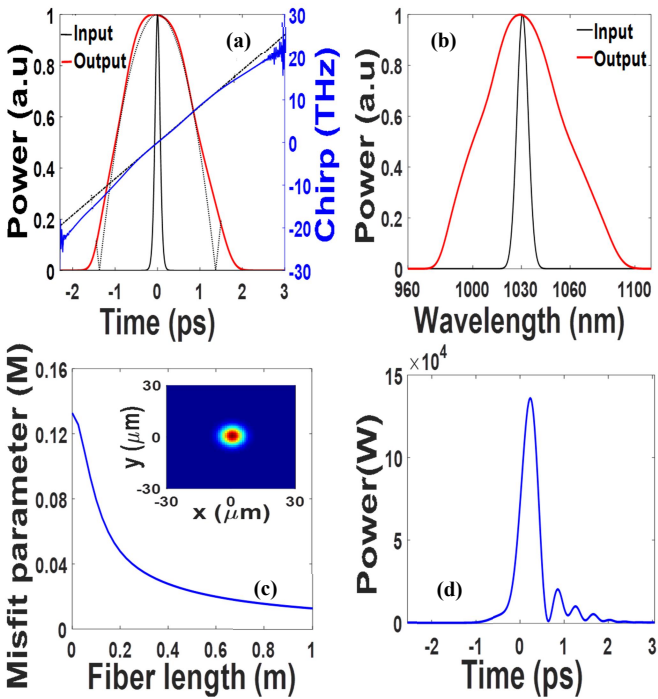

Fig.2. The temporal (a) and the spectral (b) output form of the parabolic pulse in GRIN-F, the misfit parameter (inset: the spatial beam profile of the generated pulse) and the compressed pulse (d).

Spatial beam profile evolution of the total field during the pulse propagation inside the GRIN-F is presented in the inset of Fig.2.c. The excitation of the $LP_{01}$ mode remains the Gaussian-like profile preserved along the fiber.

*The initial pulse shapes effects and influence of the initial pulse chirp*

In this part, we highlight the influence of the input pulse shapes and the initial pulse chirp parameters on the parabolic pulse generation process.
In order to understand the effects of different pulse shapes on the parabolic pulse evolution through the propagation of 1 m GRIN-F, we used two well-known pulse shapes ($10^{th}$order super-Gaussian (quasi-rectangular) and secant-hyperbolic) different that used in the previous study. The input pulses are unchirped with the same peak power and the pulse duration (see Fig. 3). Moreover, the corresponding misfit parameters are calculated according to the distance of propagation. The pulse evolution shows that the pulse evolves into the parabolic pulse shape independent of the input pulse shape and maintaining the linear chirp over the entire pulses duration in both cases. We can see from Fig.3.a that the pulse reshaping process in the case of initial super-Gaussian pulse leads to the large parabolic pulse. The broadening factor for the pulse width is larger compared to generate by Gaussian initial pulse. The fast transformation to parabolic and the broadening are due to the increase of the dispersion effect for pulses with sharper edges. However, input super-Gaussian pulse leads to parabolic pulse shape with steeper leading and trailing edges. This feature is also appeared on the corresponding generated spectrum (Fig.3.b). In the case of secant-hyperbolic pulse form, the resulting pulse shape quite differs from the parabolic one arises in the center of the pulse compared to the Gaussian pulse. Also, the broadening factor for the pulse width is quite smaller as well as the spectrum bandwidth presented in Fig.3.b. The Misfit parameters evolutions for different cases are plotted in Fig. 3.c. It can be seen that the evolution to the parabolic profile is different. In the case of the initial super-Gaussian pulse the deviation from parabolic shape is minimal (M = 0.01), whereas the maximal deviation appears in the case of the initial secant-hyperbolic pulse (M = 0.015).
We should be noted that the peak presented in the misfit curve for the super-Gaussian pulse is due to the compressed process occurred to the pulse at the first propagation distance (0.1m) in the GRIN-F before starts to broadening. At the point of compression, the peak power of the pulse increases which enhances the nonlinear effects and makes the spectrum broadening stage occurs earlier. While, in the case of the initial Gaussian pulse and the secant-hyperbolic pulse, the curves of the misfit evolutions are stable. As discussed earlier, we choose Gaussian pulse for further studies as the misfit parameter is minimal and its evolution is stable.
In the second part, the influence of the initial pulse chirp on the evolution of the parabolic pulse is studied. We demonstrated the effect of pulse chirp sign by keeping other parameters fixed. The results presented by the variation of the misfit parameter and plotted in the Fib.3.d of both positive (C < 0) and negative (C > 0) initial pulse chirps along with unchirped one (C= 0).

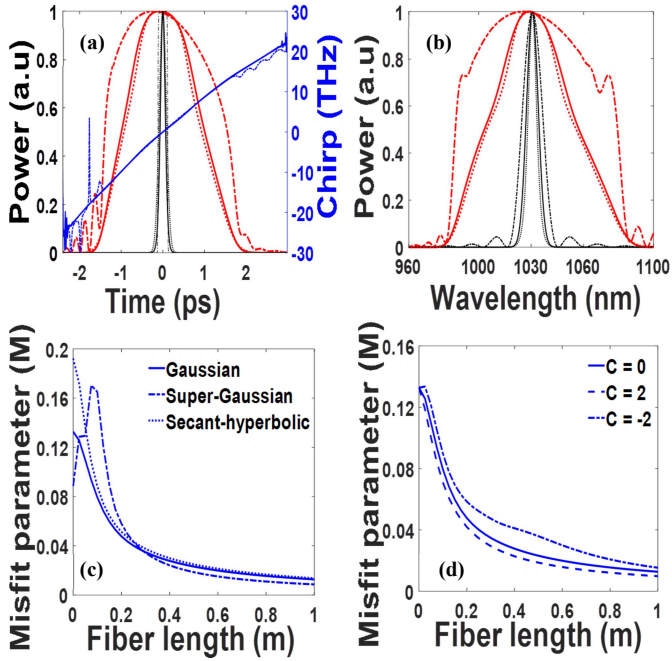

Fig.3.The temporal (a) and the spectral (b) output pulse form when the input pulse has a different initial form and the corresponding misfit parameters (c).(Gaussian: solid line, Secant-hyperbolic: dashed line and Super-Gaussian: mixed line).The misfit parameters when the input pulse has a different initial chirp (d).

The results first show that all pulse with different initial chirp sign evolve to the parabolic shape form. Significantly, the variation of the misfit parameter evolution is quite diverse depending on the initial pulse chirp parameter. For the initial negative chirp (C = -2), the pulse has occurred in the first stage a compression before starts to broadening and evolved in the parabolic shape. The initial negative chirp is changed the pulse evolution as compared to the unchirped one, when the curve of the misfit exhibit peak and deep, which indicate that the pulse is changing its chirp. The value of M is increased at the distance of 0.025 m, which correspond to the transient state propagation. The M parameter reached the minimum value of 0.016 at 1m of propagation and it is high to that one for the unchirped case (0.013). So the pulse needs to propagate much longer in the GRIN-F to reach the optimum parabolic fit. While the positive initial chirp (C = 2) accelerates the transformation to parabolic form and the broadening of the pulse width compared with unchirped case. This occurs because the normal dispersion as well as SPM produces positive chirp to the pulse, which is additive to the initial one. The M parameter reached the minimum value of 0.01 at 1 m of propagation, which indicates that the generation of the parabolic pulse is reached about 0.1 m earlier than that of the unchirped case with the same M value. We should be noted that the misfit evolutions for all chirp signs are stable at 1 m of propagation. These results demonstrate that the different initial chirp sign can be used to generate desired parabolic pulses. In addition, we also recognized that the initial pulse shapes and chirps do not disturbed the Gaussian-like spatial beam profile.

### B. Under the condition of the excitation by the equal energy among all the modes

In this case, we first initialized simulations with equal energy distribution in each of the six modes of the proposed GRIN-F. In order to avoid the strong mode coupling between the modes, the total energy of 1.5 nJ among all the modes is used and keeping the same initial pulse parameters as above. As a result of pulse propagation in 1m GRIN-F length, all the modes evolve towards the respective parabolic shapes in both the temporal and the spectral profiles with different energy as shown in Fig.4.aand b. We noted that, due to the very small walk-off for 1m of GRIN-F, all the modes are well centred in temporal and spectral domains. More interestingly though of all the six modes, only $LP_{01}$ and $LP_{02}$ modes are significantly gain energy. The difference of the energy among all the modes is due to the modal energy exchange between them as shown in Fig.4.d. The initial energy starts to transfer from the HOMs into the $LP_{01}$ and $LP_{02}$ modes at distance of 0.1 m. A steady state of energy evolution is reached after ~ 0.9 m with 50% of the energy presented in the $LP_{01}$ and $LP_{02}$ modes (20% in the $LP_{01}$ and 30% in the $LP_{02}$ mode). In the Fig.4.c, we also calculated the evolution of Misfit parameters for all the modes and we notice that the minimum value is quite diverse depending on the energy of each mode. The formation of the parabolic shape is strongly governed by the SPM and the dispersion action on the pulse and higher energy provides stronger impact of SPM and mode reshaping is relatively sufficient for example the case for $LP_{02}$ mode with M parameter of 0.039. On the other hand, in the case of the $LP_{11a}$ mode which is lost the initial energy, the dispersion dominates and mode reshaping is occurred quickly (M = 0.025). Notably, the misfit parameter increases considerably with increasing of the mode energy, which leads to increasing the deviation from the parabolic shape. Besides, the required distance to achieve a parabolic shape with good fit (M <0.04) is varied from 10-40 cm from all the modes. The distances are 0.6 m for $LP_{11a}$ mode, 0.8 m for $LP_{01}$ and $LP_{11b}$ modes, 0.8 m for $LP_{21a}$ and 1m for $LP_{02}$ and $LP_{21b}$ modes, respectively.

As a result of intermodal energy exchange and the parabolic pulses formation process, the input beam profile is evolved from a speckled pattern to a centered beam profile as shown in Fig.4.d and e, respectively.

To investigate the origin of the spatial behaviors, we then varied the initial pulse energies. Fig.5 presents the variation of the output pulses, the modal energy evolution and the output spatial beam profiles with the different initial energies. Even the initial energy distribution between the modes is equal, the final results revealed significant spatial differences entirely. At relatively low pulse energy less than 0.15 nJ (called linear propagation regime), there is no significant energy exchange between the modes. The energy in each mode maintains the initial value without providing a parabolic reshaping of the modes (Fig. 5.a). The spatial beam profile is containing all the modes with nearly the same energy (inset of Fig. 5.a). When the input pulse energy is increased, energy exchange between the modes starts to evolve and more energy transferred into the $LP_{02}$ mode is observed. The latter starts to increase its energy content (30 %) when the pulse energy reaches 1.5 nJ. For the total energy of 15 nJ case, it becomes dominant with about 50% of the total energy. The steady state of the $LP_{02}$ mode energy evolution is reached at a distance of 0.4 m. This is consistent with the observations of the parabolic pulse as it evolves toward the maximum before break-up (Fig.5.b). More interestingly, a nonlinear phenomenon called beam auto-selection is observed, when the initial multimode beam profile is evolved to $LP_{02}$ mode beam profile (inset of Fig. 5.b). On the other hand, the initial energy of 30 nJ is strong enough to cause significant strong nonlinear modes coupling. The modes undergo significant pulse evolution under energy exchange between them, which deteriorate the parabolic shape formation (Fig. 5.c). We note that in this case there is no mode predominant over others. As a result, the spatial beam profile is containing all the modes with different energy (inset of Fig. 5.c). Regarding the above results, we can conclude that the parabolic pulse formation process has a prominent effect on the spatial behaviour inside the GRIN-F. Moreover, the occurrence of beam auto-selection depends on the parabolic pulses formation process remains spatial beam profile stable for more than a decade of the initial pulse energy variation (from 1.5 to 15 nJ).

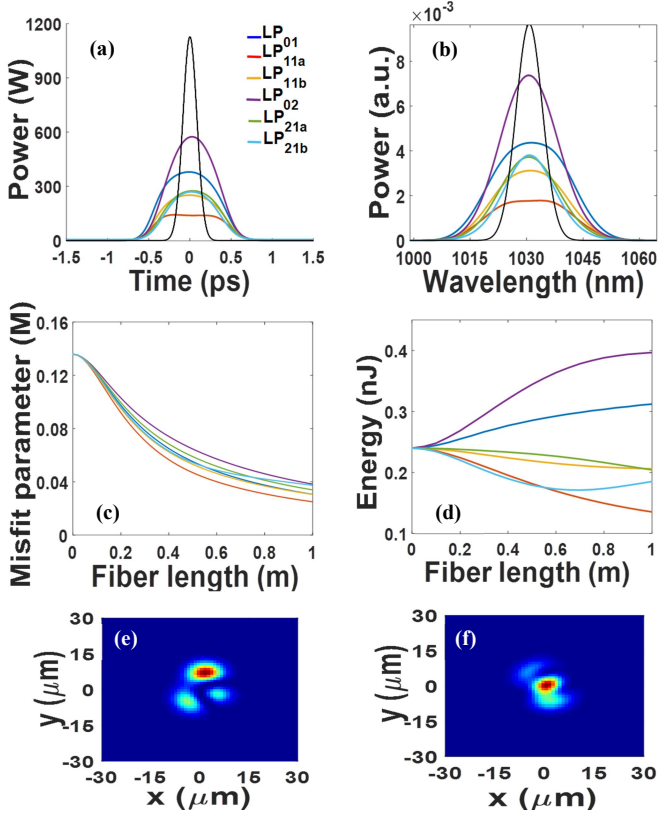

Fig.4. The temporal (a) and the spectral form of all the modes (b), the misfit parameters (c), the modal energy evolution (d) and the spatial beam profile of the initial pulse (d) and the generated pulse (e).

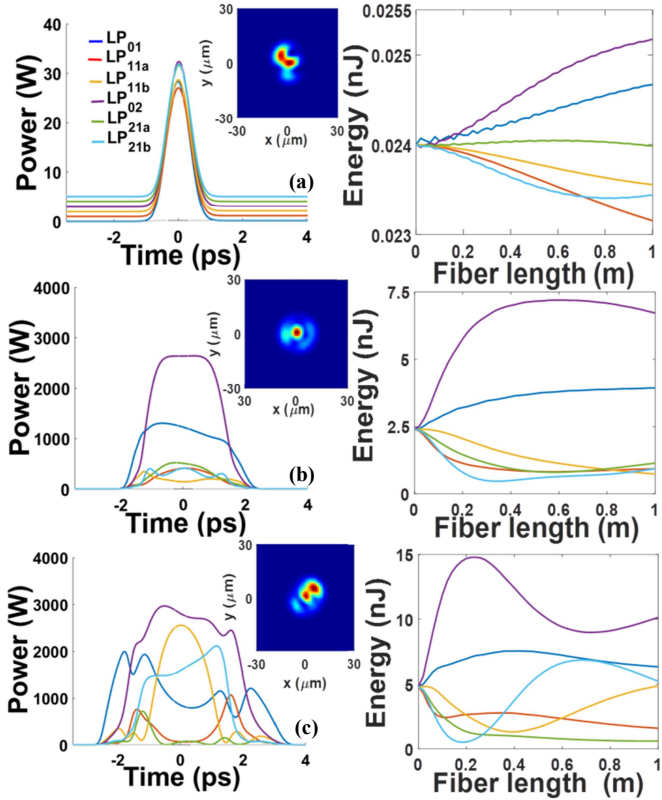

Fig.5. The output pulse (inset: output beam profile) and the modal energy evolution for initial pulse energy of 0.15 nJ (a),15 nJ (b) and 30 nJ (c).

IV. PARABOLIC PULSE GENERATION THROUGH THE PROPOSED STEP-F

A. Under the condition of the predominant fundamental mode excitation

Alike the cases of the GRIN-F, we assume Gaussian-shaped input pulse with pulse duration of 200 $fs$ propagates in 1m of STEP-F. The input pulse energy is about 50 nJ corresponding the pulse peak power is considered to be 250 kW. We first set this energy mostly coupled in the $LP_{01}$ mode as 99 %. The input pulse energy is chosen to ensuring the formation of parabolic pulse and to avoid the energy transfer into the other HOMs. The characteristic nonlinear length in this case is 20 mm. We should indicate here, that the initial parameters are chosen to give the similar dynamics proposed in the GRIN-F case. The input pulse evolution shows the characteristics of the self-similar regime where a decrease in amplitude and corresponding increase in pulse width as well as a substantial broadening of the spectrum is evident. The different input pulse forms such as Gaussian, secant-hyperbolic or super-Gaussian evolves along the STEP-F to awards a parabolic shape with linear chirp as depicted in Fig.1, 2 and 3, respectively. The only difference is the larger Gaussian-like beam size due to the different mode field diameters of $LP_{01}$ mode for both MMFs.

B. Under the condition of the excitation by the equal energy among all the modes

The STEP-F proposed in this work presents lower nonlinear coefficient than the GRIN-F. The formation of the parabolic shape for all modes is strongly governed by the SPM and the dispersion. Therefore higher initial energy among all the modes is required for this type of fiber. However, using highly energy leads to strong nonlinear modes coupling, which directly affects the multimode pulse evolution as well as the spatial beam profile. To avoid this issue, we propose a new way with weak value of initial pulse energy and increasing the

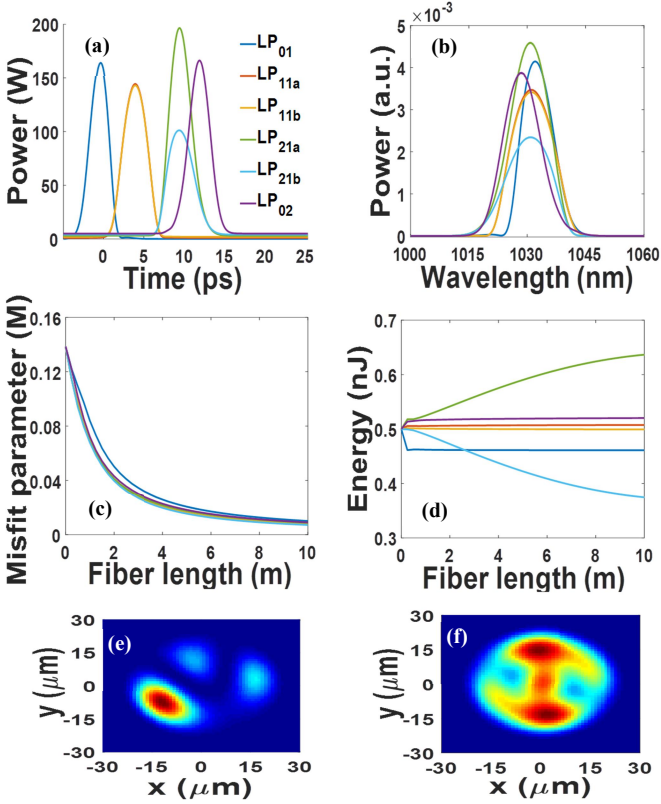

Fig.6. The temporal (a) and the spectral form of all the modes (b), the misfit parameters (c), the modal energy evolution (d) and the spatial beam profile of the initial pulse (e) and the generated pulse (f).

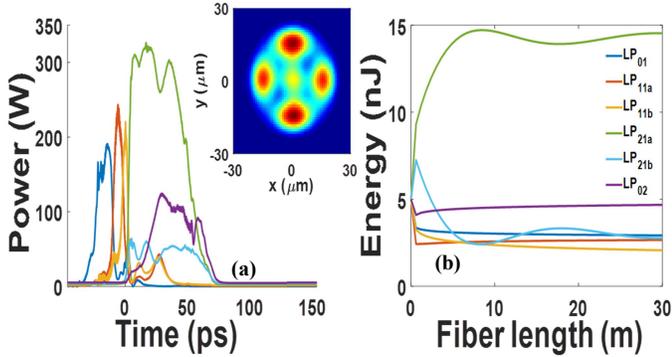

Fig. 7. The output pulse (a) and the modal energy evolutions (b). (inset: output spatial beam profile)

fiber length to permit the parabolic reshaping of all the modes during its propagation. In this case, the total energy of 3 nJ among all the modes and 10 m STEP-F length are used. As a result, all the modes evolve towards the parabolic shapes in both temporal and spectral profiles as shown in Fig.6.a and b. A weak energy exchange is only observed between the modes $LP_{21a}$ and $LP_{21b}$. $LP_{21a}$ mode is the highest and other modes are almost kept the initial energy level (see Fig.6.d). We notice a large walk-off between modes $LP_{01}$ and $LP_{02}$ due to the large modal dispersion of the STEP-F, which means that the coupling between them may be less efficient. However, $LP_{02}$ mode overlaps with the $LP_{01}$ mode at the earlier distance of propagation (0.1 m), which also caused energy transfer from $LP_{01}$ into $LP_{02}$ mode as shown in Fig. 6.c. Furthermore,

Fig.6.b shows clearly the spectral shift caused on the spectrums of these modes, confirm that, $LP_{0m}$ modes coupling is an intrinsic feature of MMF [19]. Unlike the case of the GRIN-F, the misfit parameter is evolved in the same manner among all the modes (M ~ 0.01) because all the modes propagate with a steady state of energy evolution as shown in Fig.6.d, which makes the parabolic reshaping process stable among all the modes during propagation. Notably, the small deviation observed in the $LP_{01}$ mode evolution is due to the coupling with the $LP_{02}$ mode in the first distance of propagation as mentioned above. The spatial beam evolution in the STEP-F presented in Fig.6. e and f (for the input and the output beam profile, respectively) shows that the spatial profile containing all the modes and the output pulse have the same modal components although the $LP_{21a}$ mode is the highest.

Next, we also investigate the effect of higher pulse energy in order to understand the energy exchange mechanism between modes. For total initial energy of 30 nJ among all of the modes, the output pulse, the modal energy evolutions and the output beam profile are presented in Fig.7. a and b. The pulse energy increased results the strong energy exchange between the modes with more energy transferred into the $LP_{21a}$ mode. Thus, the $LP_{21a}$ mode emerges and dominates with steady state of energy evolution up to 14.5 nJ at the output. The $LP_{02}$ mode keeps its initial energy of 5 nJ. Whereas, the other modes lost energy with nearly the same among of ~ 2-3 nJ. The corresponding spatial profile in the inset of Fig.7.a is dominated by the shape of the $LP_{21a}$ mode with more than 50% of the total energy. The modes undergo significant evolution under energy exchange between them, which deteriorate the parabolic shape formation. The spatial beam profile is shown to be of the $LP_{21a}$ beam. So, as in the GRIN-F, the parabolic pulse formation process has also a prominent effect on the spatial behaviour inside the STEP-F, which remains same spatial beam profile. The occurrence of beam auto-selection is only observed for the $LP_{21a}$ mode; however, it is dominated beyond the parabolic formation regime.

## V. CONCLUSION

In this paper, the various numerical simulations are carried out to investigate the spatiotemporal dynamics of self-similar parabolic pulse evolution in MMFs. We also demonstrated that parabolic pulse could have a prominent effect on the spatial behaviour in two different kinds of MMFs (GRIN-F and STEP-F). The results showed that the parabolic pulses can be generated and propagated self-similarlyin both MMFs depending of initial conditions. The characteristics of the parabolic pulses generated present similar behaviour as obtained in SMFs with temporal and spectral parabolic shape and linear chirp in both MMFs. Compressions of the output pulses lead to the generation of 40 fs pulses. We have furthermore studied the impact of various initial pulse shapes and chirp parameters on the pulse reshaping process. It was found that the initial pulse parameters can be used to generating the tailoring parabolic pulses. The transverse distribution intensity of the beams remains Gaussian-like profile along the both MMFs when the most energy is coupled in the fundamental mode. In the second case, when the both MMFs excited in a way to have equal energy distribution among all the modes, we have reported the ways in whether all the modes converge to the parabolic shapes. Furthermore, we have studied the effect of the initial energy of the input pulse on the resulting parabolic pulse and the corresponding spatial beam in both fibers. We have also shown that the parabolic pulse generation have a prominent effect on the spatial behavior, thus the beam profile, of output pulses, which could be different for same MMF with different initial pulse energy. Moreover, a spatiotemporal nonlinear dynamic, passive auto-selection of one mode is investigated. Parabolic pulse formation process plays critical roles in these dynamics. From the above results, we anticipate that the spatiotemporal parabolic pulses are promising to be exploited as convenient platforms for providing a simple framework for the interpretation of complex nonlinear dynamics in MMFs.